\documentclass[10pt,aps,prl,twocolumn,superscriptaddress]{revtex4-1}

\usepackage{epsfig,latexsym,cancel,amssymb,amsmath,verbatim,mathrsfs}
\usepackage{color}
\usepackage{graphicx}
\usepackage{hyperref}

\newcommand{\bea}{\begin{eqnarray}}
\newcommand{\eea}{\end{eqnarray}}
\newcommand{\beq}{\begin{eqnarray}}
\newcommand{\eeq}{\end{eqnarray}}

\newcommand{\lrsf}[2]{\left[\frac{#1}{#2}\right]}

\newcommand{\gev}{\, {\rm GeV}}
\newcommand{\mev}{\, {\rm MeV}}
\newcommand{\nnmb}{\nonumber}

\begin{document}
\title{Cosmic Archaeology with Gravitational Waves from Cosmic Strings
}

\author{Yanou Cui}
\email[]{yanou.cui@ucr.edu}
\affiliation{Department of Physics and Astronomy, University of California, Riverside, CA 92521, USA}
\author{Marek Lewicki}
\email[]{marek.lewicki@fuw.edu.pl}
\affiliation{ARC Centre  of  Excellence  for Particle  Physics  at  the  Terascale (CoEPP) \& CSSM, Department of Physics, University of Adelaide, South Australia 5005, Australia}
\affiliation{Faculty of Physics, University of Warsaw ul.\ Pasteura 5, 02-093 Warsaw, Poland}
\affiliation{Kings College London, Strand, London, WC2R 2LS, United Kingdom}
\author{David E. Morrissey}
\email[]{dmorri@triumf.ca}
\affiliation{TRIUMF, 4004 Wesbrook Mall, Vancouver, BC, Canada V6T 2A3}
\author{James D. Wells}
\email[]{jwells@umich.edu}
\affiliation{Leinweber Center for Theoretical Physics, University of Michigan, Ann Arbor MI 48109, USA} 
\affiliation{Deutsches Elektronen-Synchrotron DESY, Theory Group, D-22603 Hamburg, Germany}

\begin{flushleft}
KCL-PH-TH/2017-51 
\end{flushleft}
\date{\today}

\begin{abstract}

Cosmic strings are generic cosmological predictions of many extensions 
of the Standard Model of particle physics, such as a $U(1)^\prime$ symmetry 
breaking phase transition in the early universe or remnants of superstring 
theory. Unlike other topological defects, cosmic strings can reach a scaling regime 
that maintains a small fixed fraction of the total energy density of 
the universe from a very early epoch until today. If present, they will oscillate and generate gravitational waves with a frequency spectrum that imprints the dominant sources of total
cosmic energy density throughout the history of the universe.
We demonstrate that current and future gravitational wave detectors, such as LIGO and LISA, 
could be capable of measuring the frequency spectrum of gravitational waves 
from cosmic strings and discerning the energy composition of the universe 
at times well before primordial nucleosynthesis and the cosmic 
microwave background where standard cosmology has yet to be tested. 
This work establishes a benchmark case that gravitational waves 
may provide an unprecedented, powerful tool for probing the evolutionary 
history of the very early universe. 

\end{abstract}


\maketitle

\section{Introduction\label{sec:intro}}

Gravitational waves~(GW), vibrations of spacetime itself proposed by Einstein
in 1916, were recently observed directly for the first time by the LIGO 
collaboration~\cite{TheLIGOScientific:2016wyq}.  The source of these 
signals were black hole binaries, and more recently a neutron 
star binary~\cite{TheLIGOScientific:2017qsa}.
Future measurements of such astrophysical events by the LIGO~\cite{TheLIGOScientific:2014jea} 
and Virgo~\cite{TheVirgo:2014hva} detectors 
and the proposed LISA~\cite{Audley:2017drz}, 
BBO, and DECIGO~\cite{Yagi:2011wg} detectors will usher in a new era 
of observational astronomy and a much better understanding
of the largest compact objects in the universe.

Gravitational waves may also provide a unique test 
of fundamental microphysics and early universe cosmology~\cite{Binetruy:2012ze,Arvanitaki:2016qwi,Bird:2016dcv,Croon:2017zcu}. 
For example, primordial inflation~\cite{Abbott:1984fp}, 
cosmic strings~\cite{Hindmarsh:1994re,Vilenkin:2000jqa} 
and cosmological first-order phase transitions~\cite{Weir:2017wfa} 
are all expected to create GWs.  
While the GWs from inflation are generally below the sensitivity
of current and planned future detectors~\cite{Smith:2005mm,Ananda:2006af,Guzzetti:2016mkm}, 
GWs from cosmic strings or phase transitions can produce observable signals.
In many cases, the potential sensitivity of GW detectors to these phenomena
extends well beyond the reach of other foreseeable laboratory 
and cosmological tests. 
In this \textit{Letter} we focus on the GWs from cosmic strings and demonstrate 
their unique potential for exploring cosmological history. 

Cosmic strings are stable one-dimensional objects characterized by a
tension $\mu$. 
They arise in superstring theory as fundamental or $(p,q)$ 
strings~\cite{Copeland:2003bj,Dvali:2003zj}.
They can also emerge as vortex-like solutions of field theory~\cite{Nielsen:1973cs}, such as configurations that wrap one or more times at spatial infinity
in theories with a spontaneously broken $U(1)$, in which case the tension
is related to the symmetry breaking scale by $\mu \sim \sigma^2$~\cite{Kibble:1976sj}.
In the cosmological setting, cosmic strings form 
a network of horizon-length long strings together with a collection 
of closed string loops~\cite{Hindmarsh:1994re,Vilenkin:2000jqa}.  
Such a network would distort the cosmic microwave background~(CMB), and current observations limit $G\mu < 1.1\times 10^{-7}$, where $G$ is Newton's 
constant~\cite{Charnock:2016nzm}.

Gravitational radiation is a key part of the evolution 
of a cosmic string network~\cite{Hindmarsh:1994re,Vilenkin:2000jqa}.  
Long strings intercommute to create closed string loops.  These loops then oscillate,
emitting energy in the form of gravitational waves until they decay away~\cite{Vilenkin:1981bx,Vachaspati:1984gt,Turok:1984cn,Burden:1985md}.
Together, the processes of intercommutation, oscillation, and emission
allow the string network to shed energy efficiently and prevent it from
dominating the energy density of the universe.  Instead, a cosmic string
network is expected to reach a scaling regime
in which it tracks the total energy density with fraction on the order
of $G\mu$~\cite{Albrecht:1984xv,Bennett:1987vf,Allen:1990tv}.
The scaling property of cosmic strings leads to a stochastic background of gravitational radiation built up from GW emission over the history of the string 
network~\cite{Vilenkin:1981bx,Caldwell:1991jj}. 
  
In this \emph{Letter} we show that the frequency spectrum of GWs 
from cosmic strings can be used to look back in time and test the 
evolutionary history of the universe. For each frequency band of the background observed today, the emission was dominated by strings in a particular era of the early universe~\cite{Caldwell:1991jj}. The standard thermal picture for the evolution of the cosmos is primordial 
inflation followed by reheating to a high temperature, 
and a subsequent long period in which the expansion of the universe is
driven by a dominant energy density of radiation 
until the more recent transitions to matter and then dark energy domination.
Evidence for this \textit{standard cosmology} comes primarily 
from observations of the CMB~\cite{Ade:2015xua}
and the successful predictions of Big Bang Nucleosynthesis~(BBN),
corresponding to cosmic temperatures below
$T\simeq 5\,\mev$~\cite{Cyburt:2015mya}. Measurements of the GW frequency spectrum from cosmic strings
by current and planned detectors could test the standard cosmology
at even earlier times and possibly reveal deviations from it.

To demonstrate the power of GWs from cosmic strings to probe the 
very early universe, we study the frequency spectrum emitted by
a string network in the standard cosmology and in two
well-motivated variations.  We focus on an ideal Nambu-Goto~(NG)
cosmic string network and apply the results of recent simulations of string
networks to compute the GW spectrum.
We show that 
a combination of current and planned GW
detectors with different frequency sensitivities may enable 
us to \textit{reconstruct a timeline of cosmic history well beyond the BBN epoch}.

\section{GW from Cosmic Strings\label{sec:csgw}}

  Oscillating closed string loops are typically the dominant
source of GWs from a cosmic string network in the scaling regime.
The length $\ell$ of a string loop created by the network at time $t_i$
evolves according to
\beq
\ell = \alpha t_i - \Gamma G\mu(t-t_i) \ .
\label{eq:looplength}
\eeq
The first term is the initial loop size as a fraction $\alpha$ of the 
formation time $t_i$, \textit{i.e.}, a fraction of the horizon size.
Recent cosmic string simulations find that about 10\% of the energy released by the long string network goes
to $\alpha\simeq 10^{-1}$ large loops, with the remaining $90\%$
going to the kinetic energies of highly-boosted smaller 
loops~\cite{Vanchurin:2005pa,Olum:2006ix,Martins:2005es,Ringeval:2005kr,BlancoPillado:2011dq,Blanco-Pillado:2013qja,Blanco-Pillado:2017oxo}.
The kinetic energy redshifts away and is not transferred to GWs.
The second term above describes the shortening of the loop
as it emits gravitational radiation, characterized by the dimensionless
constant $\Gamma \simeq 50$~\cite{Vilenkin:1981bx,Turok:1984cn,Quashnock:1990wv,Blanco-Pillado:2013qja,Blanco-Pillado:2017oxo}. 

  String loops emit GWs from normal mode oscillations at frequencies
$f_{emit}=2k/\ell$, $k \in \mathbb{Z}^{+}$. After emission, the frequency of the GW redshifts as $a^{-1}$, 
where $a(t)$ is the cosmological scale factor.
For a given GW frequency $f$ observed today from mode $k$, this 
implies the emission time $\tilde{t}$ is related to the loop formation time by
\beq
f = \frac{a(\tilde{t})}{a(t_0)}
\,
\frac{2k}{\alpha t_i - \Gamma G\mu(\tilde{t}-t_i)} \ ,
\label{eq:ftoday}
\eeq
where $t_0$ is the current time.
  
  The stochastic GW background depends on the rate of loop production 
by the cosmic string network.  
We model this using the velocity-dependent one-scale~(VOS) 
model~\cite{Martins:1995tg, Martins:1996jp,Martins:2000cs}, with a loop chopping efficiency of $\bar{c} = 0.23$~\cite{Martins:2000cs}.
Applying local energy conservation and considering only large loops,
this yields a loop formation rate per unit volume at time $t_i$ of
\beq
\frac{dn_{loop}}{dt_i} = (0.1)\,\frac{C_{eff}(t_i)}{\alpha}\,t_i^{-4} \ ,
\eeq 
where the first factor accounts for the $10\%$ of the network energy going to large loops discussed earlier~\cite{Blanco-Pillado:2013qja,Blanco-Pillado:2017oxo}.
The function $C_{eff}(t_i)$ depends on the redshift scaling of the dominant energy density 
$\rho$ of the universe.  When $\rho$ is dominated by a single source,
it scales approximately as
\beq
\rho ~\propto~ a^{-n} \ .
\label{eq:neff}
\eeq
For $n = 3$ (matter domination), 4 (radiation domination), 
and 6 (kination -- to be explained later),
the VOS model predicts $C_{eff} = 0.41,\,5.5,\,30$, respectively. 

  Summing over all harmonic modes, 
the GW density per unit frequency seen today is
\beq
\Omega_{GW}(f) = \frac{f}{\rho_c}\frac{d\rho_{GW}}{df} =
\sum_{k}\Omega^{(k)}_{GW}(f) \ ,
\eeq
with
\beq
\Omega^{(k)}_{GW}(f) &=&\frac{1}{\rho_c}\frac{2k}{f}\frac{(0.1)\,\Gamma_kG\mu^2}{\alpha(\alpha+\Gamma G\mu)}
\label{eq:omgw}
\\
&\times&
\int_{t_F}^{t_0}\!\!d\tilde{t}\,\;\frac{C_{eff}(t_i)}{t_i^{4}}\!
\lrsf{a(\tilde{t})}{a(t_0)}^5\!\lrsf{a(t_i)}{a(\tilde{t})}^3 \Theta(t_i-t_F) \ ,
\nnmb
\eeq
where $\rho_c=3H_0^2/8\pi G$ is the critical density, 
$\Gamma_k = \Gamma/(3.60\cdot k^{4/3})$~\cite{Blanco-Pillado:2013qja,Blanco-Pillado:2017oxo},
 $t_i$ is obtained by inverting Eq.~\eqref{eq:ftoday} and $t_F$ is the formation time of the string network.

 \begin{figure}[ttt]
 \includegraphics[height=5.5cm]{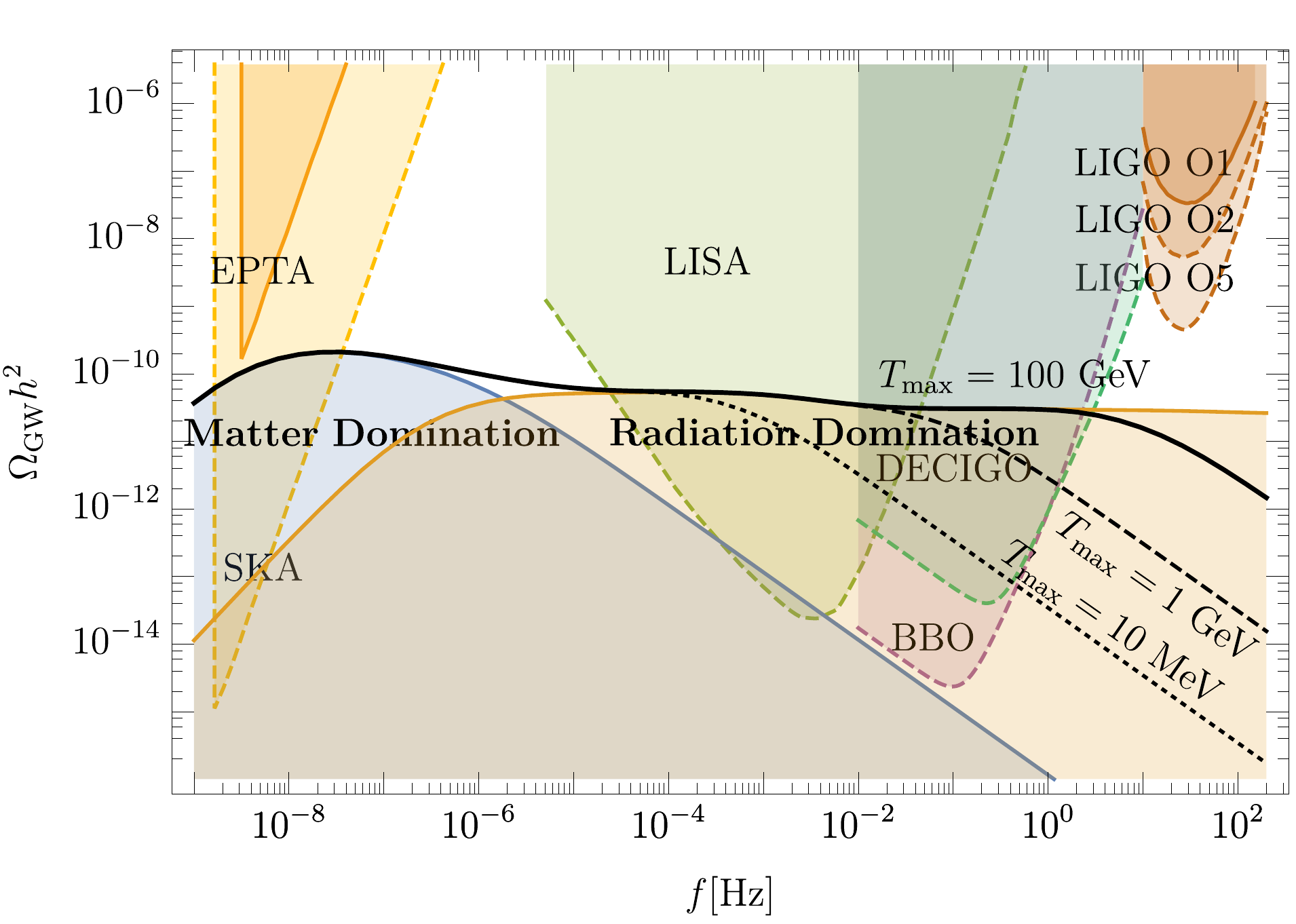}
 \caption{\label{fig:gwprod}
Frequency spectrum $\Omega_{GW}h^2$ of GWs from a cosmic string network 
with $G\mu = 5\times 10^{-12}$ and $\alpha=10^{-1}$ 
assuming a standard cosmological history.  
The blue~(orange) region indicates contributions emitted during matter~(radiation) domination with $n = 3$~($n \simeq 4$).  The falling black lines indicate the effect on the spectrum if only those loops created below the temperatures $T_{max} = 10 \,\mev$~(dotted), $T_{max} = 1\,\gev$~(dashed), and  $T_{max} = 100\,\gev$~(solid).  The upper shaded regions show current and future experimental GW sensitivities.}
 \end{figure}

  In Fig.~\ref{fig:gwprod} we show 
the frequency spectrum of GWs $\Omega_{GW}h^2$ from cosmic strings
for $G\mu = 5\times 10^{-12}$, 
$\alpha=10^{-1}$, and a standard cosmological history. 
Also shown are the current sensitivity bands 
of LIGO~\cite{TheLIGOScientific:2014jea,TheLIGOScientific:2016wyq,Thrane:2013oya}, and the projected sensitivities of LISA~\cite{Bartolo:2016ami}, 
DECIGO, and BBO~\cite{Yagi:2011wg}.
The upper left solid triangle indicates the current limit 
from timing measurements 
by the European Pulsar Timing Array~(EPTA)~\cite{vanHaasteren:2011ni}, 
with the expected sensitivity of the future Square Kilometre 
Array~(SKA)~\cite{Janssen:2014dka} shown below.  The EPTA
limit implies $G\mu \lesssim 10^{-11}$ giving the
strongest current bound on the cosmic string network. 
Our results are consistent with the recent calculations 
of Refs.~\cite{Blanco-Pillado:2017rnf,Ringeval:2017eww}.

Fig.~\ref{fig:gwprod} illustrates the key relationship between
the GW frequency spectrum and the cosmological era at which a given
frequency today was produced.  
At higher frequencies, the result of Eq.~\eqref{eq:omgw}
is approximately independent of $\alpha$ and scales as
\begin{equation}\label{eq:gwspec}
\Omega_{GW}(f)\propto
\begin{cases}
 f^{\frac{8-2n}{2-n}} \quad n> 10/3 \\
 f^{-1} \quad \quad n\leq 10/3
 \end{cases}
\end{equation}
This yields a flat spectrum for radiation and a $1/f$ spectrum for matter.
At lower frequencies, the scaling (for $\alpha \gg \Gamma G\mu$) goes
like $f^{3/2}$.
The characteristic flatness of the spectrum at higher frequencies from
early radiation domination implies that deviations from this scenario
would be \textit{dramatic}.

\section{Tests of Non-Standard Cosmologies\label{sec:mod}}

Cosmic string scaling implies that the relic spectrum of GWs today originated
from an extended period of evolution of the early universe.
As a result, the GW spectrum from cosmic strings can 
probe
deviations from the standard cosmological evolution and identify the specific
era in which it occurred.  
To illustrate this,
we focus on two well-motivated non-standard histories.
The first is a transient period of matter domination prior to the standard 
radiation era, corresponding to $n=3$ in Eq.~\eqref{eq:neff}.
This can arise from a large density of a long-lived massive particle
or from the oscillation of a scalar moduli field in a quadratic 
potential~\cite{Moroi:1999zb}.  Matter dominance typically ends during a 
reheating phase in which the relevant species decays to light 
Standard Model~(SM) particles.  
The second non-standard cosmology we consider is a period of ``kination'',
with $n > 4$ in Eq.~\eqref{eq:neff}.  This can arise from the 
oscillation of a scalar field in a non-quadratic potential:
for $V(\phi) \propto \phi^{N}$ one obtains $n = 6N/(N+2)$,
which can occur in quintessence models for dark energy 
or inflation~\cite{Salati:2002md,Chung:2007vz}. 

For both non-standard scenarios, above,
let $t_{\Delta}$ be the time at which the universe transitions 
to the standard period of radiation domination. The evolution of the energy density of the universe during and after 
the non-standard phase can be parameterized according to
\beq
\rho(t) =
\begin{cases}
\rho_{st}(t_{\Delta})\,\left[\frac{a(t_{\Delta})}{a(t)}\right]^{n}
 & \ {;} \ t < t_{\Delta}
\\
\rho_{st}(t) & \ {;} \ t \geq t_{\Delta}
\end{cases}
\eeq
where $\rho_{st}$ is the energy density extrapolated assuming the standard
cosmological history, and $n=6$~($n=3$) for early kination~(matter) domination. 
We also define $T_{\Delta}$
as the
temperature at time $t_{\Delta}$ when radiation domination 
resumes.  In scenarios with early matter domination, 
$T_{\Delta}$ coincides with the reheating temperature.  
For both the matter-dominated and kination scenarios, $T_{\Delta} \gtrsim 5\,\mev$ 
is needed for consistency with BBN~\cite{Hannestad:2004px}.

\begin{figure}[ttt]
 \includegraphics[height=5.5cm]{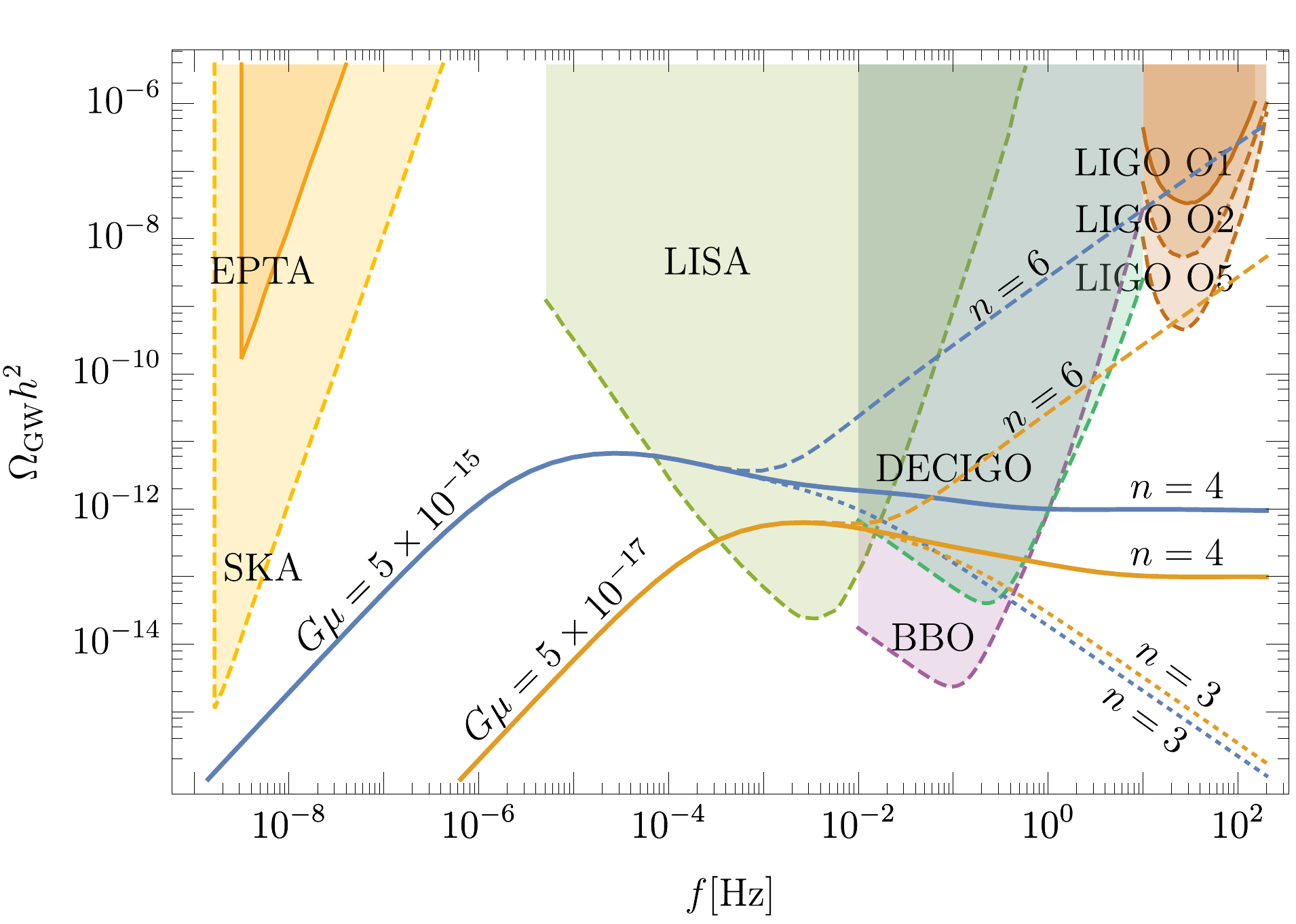}
 \caption{\label{fig:gwmod}
Gravitational wave frequency spectra $\Omega_{GW}h^2(f)$ from a cosmic string 
network with $G\mu = 5\times 10^{-15}{\rm{(blue)}},\,5\times 10^{-17}${\rm{(orange)}} 
and $\alpha=10^{-1}$ for standard and non-standard 
cosmological histories.  The solid lines show the spectra 
for the standard evolution while the dashed~(dotted) lines 
show the GW spectra for an early $n=6$ kination~($n=3$ matter)
era ending at temperature $T_{\Delta} = 5\,\mev$.
The upper shaded regions indicate the current and future sensitivities 
of GW detectors and pulsar timing arrays.} 
 \end{figure}

In Fig.~\ref{fig:gwmod} we illustrate the effect of non-standard cosmologies
on the frequency spectrum of GWs from a cosmic string network
with $G\mu= 5\times 10^{-15},\,5\times 10^{-17}$ and $T_{\Delta} = 5\,\mev$,
along with the experimental sensitivities 
current and future detectors
as in Fig.~\ref{fig:gwprod}.
The deviations in the GW spectra shown Fig.~\ref{fig:gwmod} due to
non-standard cosmologies are 
dramatic and potentially observable at future
GW detectors. The effect of an early kination phase is especially
distinctive, and we find current LIGO data already gives a limit 
of $G\mu\lesssim 5\times 10^{-15}$, orders of magnitude smaller than other bounds on the string network. 
The limits are less stringent if kination ends earlier 
than $T_\Delta=5\, {\rm MeV}$; nevertheless, the effect illustrates 
the power of GWs from cosmic strings to probe the early universe.
In contrast,
an early phase of matter domination tends to suppress the GW spectrum 
at high frequencies, putting it out of range of LIGO detection or constraint. 
However, the turn-over it induces is still potentially observable 
by future detectors such as LISA, DECIGO and BBO.

\begin{figure}[ttt]
 \includegraphics[height=5.5cm]{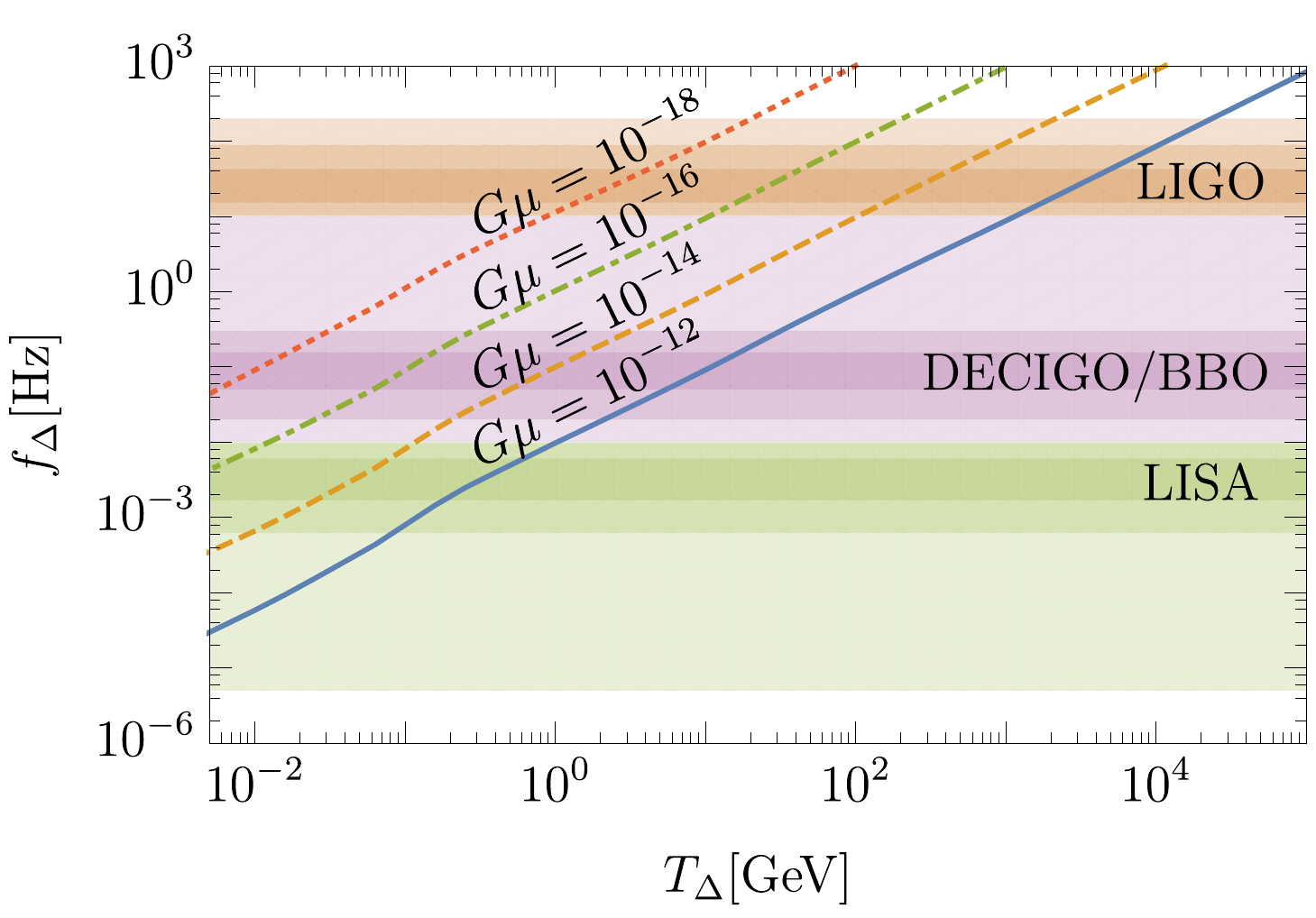}
 \caption{\label{fig:fdelta}
Frequency $f_{\Delta}$ at which the GW spectrum from cosmic strings 
is altered by a non-standard cosmology as a function of $T_\Delta$ 
for several values $G\mu$ and $\alpha=10^{-1}$.  The shaded bands
indicate the sensitivities of current and future GW detectors,
with the darker regions showing the peak sensitivities.
}
 \end{figure}

An additional constraint not yet included comes from
the total radiation density of GWs, which  must not exceed the limits from 
the CMB and BBN.  
This translates into the bound~\cite{Henrot-Versille:2014jua,Smith:2006nka}
\begin{equation}
\int\!
d(\ln f)\;
\Omega_{\rm GW} \lesssim 3.8 \times 10^{-6} \ .
\label{eq:toorad}
\end{equation}
For standard early radiation domination, 
this places a moderate constraint on $\Omega_{GW}$ with a logarithmic sensitivity
to the highest frequencies created in this era.  
The bound becomes more severe for $n> 4$ since the relic GW spectrum 
now increases with frequency as a power law, Eq.~\eqref{eq:gwspec}.
This growth is expected to be cut off at a high frequency that corresponds
to the onset of the $n> 4$ phase or the creation of the string network. 
In the early $n=6$ scenario with $T_{\Delta}=5\,\mev$, the maximal temperature
in this phase is $T \simeq 4\,\gev$~($20\,\gev$) 
for $G\mu=5\times 10^{15}$~($5\times 10^{-17}$).
 As can be seen in Fig.~\ref{fig:gwmod}, the deviation in the GW spectrum
due to a non-standard cosmic history occurs at a characteristic frequency
$f_{\Delta}$ that depends on the string network parameters
and $T_{\Delta}$, but is nearly independent of the energy redshift exponent $n$. 
In Fig.~\ref{fig:fdelta} we show $f_{\Delta}$ as a function of $T_{\Delta}$
for several values of $G\mu$.  These curves follow the approximate
relation $f_\Delta \propto T_{\Delta} \left(G\mu\right)^{-\frac{1}{2}}\alpha^{-\frac{1}{2}}$ (valid for $\alpha \gg \Gamma G\mu$), which we derive in a future work \cite{GW_prep}.  
Roughly speaking, $T_{\Delta}$ characterizes the formation time of 
loops that produce the dominant contribution to GWs with frequency 
$f_{\Delta}$ today.
Figure~\ref{fig:fdelta} shows that the frequency range of LIGO could be
sensitive to non-standard cosmologies all the way back 
to $T_{\Delta} \sim 10^4\,\gev$ for $G\mu = 10^{-12}$, 
well beyond the reach of other known probes.

\section{Discussion\label{sec:disc}}
  If cosmic strings are realized in nature, they could provide a unique and powerful tool for probing the history of the early universe.  We have demonstrated that the frequency spectrum of GWs emitted by a cosmic string network depends dramatically on the energy content of the universe when they are produced.  Current and planned experiments will have the 
potential to measure such a spectrum and thereby test the evolution of the cosmos at much earlier times than ever before.  

  The reach of this method for looking back in time depends on the frequency sensitivity of GW detectors and the properties 
of the cosmic string network.  
In general, deviations from the standard cosmological evolution 
at earlier times imprint themselves on the GW spectrum at higher frequencies, 
as shown in Fig.~\ref{fig:fdelta}.  
These considerations provide strong motivation to explore new methods 
to extend the sensitivity of future GW observatories to higher frequencies 
beyond LIGO~\cite{Arvanitaki:2012cn, Chou:2015sle}.

  The frequency spectrum of GWs also depends on the distribution of
string loop sizes.  Based on cosmic string simulations~\cite{Blanco-Pillado:2013qja,Blanco-Pillado:2017oxo}, we have modeled
this by assuming that $10\%$ of the energy shed by a scaling string
network goes to loops with a fixed initial loop size parameter of
$\alpha=10^{-1}$.  Our simple prescription reproduces the radiation-era
loop length distribution found in the simulation of 
Ref.~\cite{BlancoPillado:2011dq,Blanco-Pillado:2013qja} 
and is similar to their result for the matter era, but it is also
optimistic.  Smaller values of $\alpha$ would push the spectral features 
of non-standard cosmologies to higher frequencies and potentially 
outside the range of GW detectors.  
Let us also mention that we do not know of any simulations of loop formation or scaling during a kination phase, which we have assumed in our calculation.

   In addition to the two well-motivated non-standard cosmologies
studied in this \emph{Letter}, there are many other cosmological scenarios 
that could be probed by the GW spectrum from cosmic strings.  
Our results could also be modified in theories with more complex cosmic 
string dynamics such as $(p,q)$ strings with small 
intercommutation probabilities~\cite{Jackson:2004zg} or in scenarios 
where strings emit non-gravitational radiation~\cite{Srednicki:1986xg,Vilenkin:1986zz,Damour:1996pv,Cui:2007js, Cui:2008bd, Long:2014mxa}.
We defer the investigation of such possibilities to a future work~\cite{GW_prep}.

 An important further question is whether there are other sources 
that could mimic the changes in the frequency spectrum from cosmic string 
GWs due to non-standard cosmologies.  
While it is difficult to provide a definitive answer,
we note that the flat part of the cosmic string GW spectrum is very distinctive
suggesting that it would be challenging to reproduce with a single 
alternative source. 
 
In this \emph{Letter} we have investigated the effects of non-standard cosmological histories on the frequency spectrum of stochastic GWs produced by cosmic strings.  We have demonstrated that cosmic string GWs could provide an unprecedented window on the evolution of the very early universe prior to BBN and the CMB. This work may serve as an inspiring benchmark for exploiting the full potential of GW as a new tool for probing particle physics and cosmology beyond the horizon of our current knowledge.

\begin{acknowledgments}
\noindent {\bf{Acknowledgments:}}
We would like to thank Masha Baryakhtar and Mairi Sakellariadou for interesting discussions.
The work of ML was supported by the ARC Centre of Excellence for Particle
Physics at the Terascale (CoEPP) (CE110001104), the Centre for the Subatomic
Structure of Matter (CSSM), the Polish MNiSW grant IP2015 043174 and STFC grant number ST/L000326/1. 
DM is supported by a Discovery Grant from the Natural Sciences and Engineering Research Council of Canada~(NSERC), and TRIUMF, which receives federal funding via a contribution agreement with the National Research Council of Canada~(NRC). JDW was supported in part by the DOE grant DE-SC0007859 and the Humboldt Research Award of the Humboldt Foundation.
\end{acknowledgments}

\bibliography{gwstring}

\end{document}